\documentclass[fleqn,usenatbib]{mnras}
\usepackage{graphicx}
\usepackage{amssymb}
\usepackage{amsmath}
\usepackage{mathabx}
\usepackage{txfonts}
\usepackage[T1]{fontenc}
\usepackage{ae,aecompl}
\usepackage{cleveref}
\usepackage[super]{nth}

\usepackage[dvipsnames]{xcolor}

\title[High-order SD method with Viscosity]
{Spectral Difference Method with a Posteriori Limiting: III- Navier-Stokes Equations with Arbitrary High-Order Accuracy}

\author[D. A. Velasco-Romero and R. Teyssier]{%
David A. Velasco Romero$^{1,2}$\thanks{E-mail: david.velasco@ics.uzh.ch}, Romain Teyssier$^{2}$\\
$^{1}$School of Natural Sciences, Institute for Advanced Study, NJ 08540, USA.\\  
$^{2}$Department of Astrophysical Sciences, Princeton University,
NJ 08544, USA.
}

\date{Accepted XXX. Received YYY; in original form ZZZ}

\pubyear{2024}

\begin{document}
\label{firstpage}
\pagerange{\pageref{firstpage}--\pageref{lastpage}}
\maketitle

\begin{abstract}
We incorporate an arbitrarily high-order method for the Laplacian operator into the Spectral Difference method (SD). The resulting method is capable of capturing shocks thanks to its a-posteriori limiting methodology, and therefore it is able to survive scenarios in which the dissipative scales (viscous and diffusive) are not properly described. Moreover, it is capable of capturing these scales at lower resolution compared to lower-order methods and therefore attains convergence at lower resolution. We show that the method at hand has exponential convergence when describing smooth solutions and is able to recover a high-order solution when solving the dissipative scales. 
\end{abstract}

\begin{keywords}
hydrodynamics -- methods: numerical
\end{keywords}

\defcitealias{velasco2023spectral}{Paper~I}
\newcommand\paperI{\citetalias{velasco2023spectral}}
\newcommand\paperIp{\citepalias{velasco2023spectral}}

\defcitealias{velasco2024spectral}{Paper~II}
\newcommand\paperII{\citetalias{velasco2024spectral}}
\newcommand\paperIIp{\citepalias{velasco2024spectral}}


\section{Introduction}
\label{sec:introduction}
 The Spectral Difference (SD) method has been introduced \citep{Liu2004, Liu2006} as an interesting alternative to the Discontinuous Galerkin method within the context of Finite Element (FE) methods for hyperbolic systems of Partial Differential Equations (PDE). Compared to DG, SD allows a larger time step by roughly a factor of 2, while preserving stability and arbitrary high-order accuracy.\\
 
 In the context of the recently introduced a posteriori limiting to prevent oscillatory solutions in presence of discontinuities \citep{velasco2023spectral}, SD presents a competitive advantage with respect to DG, owing to its strict equivalence to a Finite Volume (FV) update using sub-element control volumes \citep{velasco2023spectral}, without requiring the costly projection step of the DG polynomial solution to an underlying sub-element Cartesian mesh \citep{Dumbser2014, Vilar2019,vilar2024}. \\

Historically, SD has also been developed for the Navier-Stokes equations \citep{may2006spectral}, following up the original implementation for the Euler equations \citep{may2006spectral, wang2007spectral}. In the context of high-order FE methods, using explicit viscosity with a strong-enough diffusion coefficient allows one to preserve the smooth character of the solution and avoid using complex slope limiters. Explicit viscosity also provides a faster convergence of the numerical solution to a well defined ``exact'' solution that can be compared between codes at various resolution. If, on the other hand, the magnitude of the diffusion coefficient is not high-enough, the solution becomes oscillatory with positivity violations, hence the need for slope limiters \citep{van1977towards}.  Our recent implementation of the SD method \citep[][hereafter Paper~I]{velasco2023spectral} is based on a posteriori limiting to prevent oscillations and preserve positivity. The method has been successfully tested in the context of the Euler equations, for its shock-capturing capability in the case of high Mach numbers \paperIp, as well as for subsonic turbulent flows in the low Mach number regime \citep[][hereafter Paper~II]{velasco2024spectral}.

In these previous papers, we showed that the high-order SD solution was capable of reproducing the small scale non-linear features obtained by a standard second-order scheme at much higher resolution. Using statistical observables such as power spectra, we could show that higher order methods were converging faster to the ``true'' solution than low order ones. It is unfortunately very difficult to truly quantify the convergence of these methods as turbulence at high Reynolds number (or low viscosity) is a complex, non-linear process with chaotic properties. The Navier-Stokes equations, on the other hand, allow one to resolve the dissipative scale at high enough resolution and obtain a truly converged solution. 

This is precisely what motivated us to implement in this work at arbitrary high-order kinematic viscosity for the SD method and to test the performance of the method (stability, accuracy, convergence) when applied to complex turbulent flows. In \autoref{sec:method} we describe the method at hand, presenting a summary of the method described in \paperI~and \paperII~as well as the algorithm for the diffusive operator, which is the main original contribution of this work. In \autoref{sec:results} we present the numerical results of the method in viscous scenarios, in \autoref{sec:discussion} we discuss these results, and in \autoref{sec:conclusions} we draw our conclusions.

\section{Numerical method}
\label{sec:method}
In this paper, we build upon our previous work (\paperI~and \paperII) by incorporating an arbitrarily high-order method for the vscosity operator into the SD framework. Our key contributions are: (1) a detailed description of the implementation of this high-order viscosity treatment, (2) a demonstration of its effectiveness in capturing the diffusive processes at small scales at lower resolutions compared to lower order methods, (3) an analysis of the method's convergence properties for both smooth and turbulent flows, and (4) a comparison with existing high-order methods in the literature.
\subsection{Governing equations}
\label{sec:ge}
The Navier-Stokes equations (including both kinematic viscosity and thermal conduction) can be written as \citep[see e.g.][]{Mihalas84}:
\begin{equation}
    \partial_t \rho + \nabla \cdot (\rho \bold{v})=0,
\end{equation}
\begin{equation}
    \partial_t (\rho\bold{v}) + \nabla \cdot (P \bold{I} + \rho \bold{v} \otimes \bold{v}) = - \nabla \cdot \bold{\Pi},
\end{equation}
\begin{equation}
    \partial_t E + \nabla \cdot [(E+P) \bold{v}] = - \nabla \cdot (\bold{v}\cdot\bold{\Pi}) - \nabla \cdot(\chi\rho\nabla T),
\end{equation}
where $\rho$ is the density of the fluid, $\mathbf{v} =(v_x,v_y)$ is the velocity field, $E = e + \frac{1}{2}\rho v^2$ the total fluid energy density, and $P$ the pressure, $\bold{I}$ is the identity tensor, $\chi$ is the thermal diffusivity and
\begin{equation}
    \bold{\Pi} = -\nu \rho \left( \nabla\bold{v} +(\nabla\bold{v})^T - \frac{2}{3}\bold{I}\nabla \cdot \bold{v}\right)
\end{equation}
is the viscous stress tensor with viscosity coefficient $\nu$. The system of equations is closed with the equation of state of an ideal gas $P = (\gamma-1)e$, where $\gamma$ is the adiabatic index.

A passive scalar $c$ is added to represent the dye concentration of the fluid. The evolution of this quantity is governed by the following conservation low:

\begin{equation}
    \partial_t (\rho c) + \nabla \cdot (\rho c \bold{v} )= -\rho \nu_\text{dye} \nabla c
\end{equation}
where $\nu_\text{dye}$ is the diffusion coefficient for this quantity.

\subsection{Spectral Difference Method}

We first summarize the method described in \citetalias{velasco2023spectral}.
The computational domain is decomposed into non-overlapping elements, explaining why SD is a FE method. Inside each element, a continuous high-order numerical solution $\mathbf{U}(x)$ is given by:
\begin{equation}
    \mathbf{U}(x) = \sum_{m=0}^p \mathbf{U}(x^s_{a,m})\ell^s_m(x),
    \label{eq:sol}
\end{equation}
where $\{\ell^s_{m}(x)\}_{m=0}^p$ represents the set of Lagrange interpolation polynomials up to degree $p$ built on the set of $p+1$ solution points $\mathcal{S}^s=\{x^s_{a,m}\}_{m=0}^p$. The elements are indexed using the index $a$. A second set of Lagrange polynomials up to degree $p+1$ on a set of $p+2$ flux points $\mathcal{S}^f=\{x^f_{a,m}\}_{m=0}^{p+1}$ in element $a$, is used to represent the high-order approximation of the flux.
\begin{equation}
\mathbf{F}(x) =  \sum_{m=0}^{p+1} \mathbf{F}(x^f_{a,m}) \ell^f_m(x).
\label{eq:flux}
\end{equation}
We define the numerical fluxes as follows:
\begin{equation}
\hat{\mathbf{F}}(x) =  \sum_{m=0}^{p+1} \hat{\mathbf{F}}(\mathbf{U}(x^f_{a,m})) \ell^f_{m}(x),
\end{equation}
where $\hat{\mathbf{F}}(\cdot)$ denotes the numerical flux resulting from solving the Riemann problem at the interface between elements. 
The solution is then updated as:
\begin{align}
\label{eq:semi-discrete}
\frac{{\rm d}}{{\rm d}t} \mathbf{U}(x^s_{a,m},t) = {\cal L}_{a,m}(\mathbf{U})= -\sum_{m'=0}^{p+1} \hat{\mathbf{F}}(\mathbf{U}(x^f_{a,m'},t)) \ell^{\prime f}_{m'}(x^s_{a,m}),
\end{align}
where $\ell^{\prime}$ is the derivative of the Lagrange polynomials.\\

The fully discrete method is achieved when combining the SD spatial discretization with a time integration discretization. In this work, we pair SD with the so-called ADER time-integrator, so that the resulting fully discrete method can be written as:
\begin{equation}
\mathbf{U}(t + \Delta t) = \mathbf{U}(t) + \Delta t \sum_{k=0}^p w_{k} {\cal L}(\mathbf{U}^{p}_k),
\label{eq:sd_update}
\end{equation}
where $k$ iterates over the $p+1$ time slices of the ADER time-integrator (for more details, see \paperI).\\

As shown in \paperI, the SD method is exactly equivalent to the FV method applied to sub-element control volumes, allowing the fully discrete scheme to be also written as:
\begin{equation}
\bar{\mathbf{U}}_{a,m}(t + \Delta t) = \bar{\mathbf{U}}_{a,m}(t) - \Delta t \sum_{k=0}^p w_k\frac{\left(\hat{\mathbf{F}}^{f,k}_{a,m+1}-\hat{\mathbf{F}}^{f,k}_{a,m}\right)}{h_{m}},
\end{equation}
where $\bar{\mathbf{U}}_{a,m}$ is the solution averaged over the control volume, and $h_m$ is the size of the control volume $m$ within each element.

\subsection{Interpolation}
\label{sec:interpolation}
The interpolation from solution points to flux points can be performed through a Lagrange interpolation matrix:
\begin{equation}
    \mathbf{U}(x^f) = L^{fs} \mathbf{U}(x^s),
\end{equation}
where $L^{fs}$ is a matrix of $(p+2)\times(p+1)$ elements, where each element is given by:
\begin{equation}
    L^{fs}_{i,j} = \ell^s_i(x^f_j).
\end{equation}
Like wise, a reciprocal interpolation matrix can be built to move from flux points to solution points:
\begin{equation}
    L^{sf}_{i,j} = \ell^f_i(x^s_j),
\end{equation}
such that:
\begin{equation}
    \mathbf{U}(x^s) = L^{sf} \mathbf{U}(x^f).
\end{equation}
The same principle can be used to built a Lagrangian matrix to compute the gradient from flux points to solution points:
\begin{equation}
    \nabla\mathbf{F}(x^s) = \nabla(\mathbf{F}(x^f)) = L'^{sf} \mathbf{F}(x^f),
\end{equation}
where $L'^{sf}_{i,j} = \ell'^f_i(x^s_j)$.

\subsection{A Posteriori Limiting}

As described in \paperI, we use a second-order Godunov scheme \citep{1979JCoPh..32..101V} as a fallback scheme when either spurious oscillations or non-physical values are observed in the high-order candidate solution. For this, we identify ``troubled cells'' for each control volume.
The control volume fluxes pertaining to the troubled cells are recomputed with the MUSCL-Hancock scheme, hereafter FV2. We then recompute, at each time stage, the update of all subcells that share these recomputed fluxes: 
\begin{equation}
    \bar{\mathbf{U}}^{k+1}_{i} = \bar{\mathbf{U}}^k_{i} - {\left(\frac{\hat{\mathbf{F}}^{k}_{i+1/2}-\hat{\mathbf{F}}^{k}_{i-1/2}}{h_i}\right)}w_k\Delta t,
\end{equation}
where $\hat{\mathbf{F}}$ represents a combination of high-order and fall-back scheme fluxes as described in \citet{2021JCoPh.42609935H} and \citet{vilar2024}. We blend the SD and FV2 fluxes at trouble-adjacent subcells in a convex combination:
\begin{equation}
    \hat{\mathbf{F}}_{i\pm\frac{1}{2}} = (1-\theta_{i\pm\frac{1}{2}})\mathbf{F}^{\text{SD}}_{i\pm\frac{1}{2}} + \theta_{i\pm\frac{1}{2},j} \mathbf{F}^{\text{FV}}_{i\pm\frac{1}{2}}
\end{equation}
where $\theta_{i-\frac{1}{2}}= \max(\theta_{i-1},\theta_{i})$ is the blending coefficient.
This coefficient has values:
$\theta_{i}=\{1,3/4,1/4,0\}$ in troubled sub-cells, in first, and second neighbors, and elsewhere, respectively:
\begin{equation}
    \theta_{i} = 
    \begin{cases}
     1 \quad & (i) \\
     3/4 \quad &(i\pm1)\\
     1/4 \quad &(i\pm2) \\
     0 \quad &\text{otherwise}
    \end{cases}
\end{equation} 
In the 2-dimensional case the blending coefficient becomes:
\begin{equation}
    \theta_{i,j} = 
    \begin{cases}
     1 \quad & (i,j) \\
     3/4 \quad &(i\pm1,j), (i,j\pm1)\\
     1/2 \quad &(i\pm1,j\pm1) \\
     1/4 \quad &(i\pm2,j), (i\pm2,j\pm1),(i\pm2,j\pm2),\\ &(i\pm1,j\pm2), (i,j\pm2) \\
     0 \quad &\text{otherwise},
    \end{cases}
\end{equation} 
as depicted in Fig.1 of \paperII.
\subsubsection{Limiting criteria}

The limiting criteria, used to determine troubled sub-cells, consist of a Numerical Admissibility Detection (NAD) and a Physical Admissibility Detection (PAD), as described in \paperI~and similar to  \citet{Vilar2019}.\\

The NAD criteria requires for the candidate solution to be bounded by the local extrema at the previous time-step:
\begin{equation}
  \min(\bar{\mathbf{U}}_{i-1}^{k},\bar{\mathbf{U}}_{i}^{k},\bar{\mathbf{U}}_{i+1}^{k}) \leq \bar{\mathbf{U}}_{i}^{k+1} \leq \max(\bar{\mathbf{U}}_{i-1}^{k},\bar{\mathbf{U}}_{i}^{k},\bar{\mathbf{U}}_{i+1}^{k}),  
  \label{eq:nad}
\end{equation}
These criteria results to stringent, as it cannot distinguish between discontinuous and smooth extrema, where smooth extrema are considered features of the solution rather than spurious artifacts of the method. In order to relax these criteria and allow smooth extrema to evolve unlimited, we add a Smooth Extrema Detection (SED), where at least the first numerical derivative of the candidate solution must be continuous (\paperI). In this work we apply the NAD criteria on density, pressure and $c$ only, and we use a tolerance on the detection:
\begin{equation}
  \bar{\mathbf{U}}_{i,\min}^{k}-\varepsilon|\bar{\mathbf{U}}_{i,\min}^{k}| \leq \bar{\mathbf{U}}_{i}^{k+1} \leq \bar{\mathbf{U}}_{i,\max}^{k}+\varepsilon|\bar{\mathbf{U}}_{i,\max}^{k}|,  
\end{equation}
where we make the choice of $\epsilon=10^{-5}$, which has provided robust results in previous works (\paperI,\paperII).\\

The PAD criterion, requires the candidate density and pressure to be above a minimum value. For this work, we use $\rho_{\min}=10^{-10}$ and $P_{\min}=10^{-10}$.

\subsection{Diffusion operators}

The implementation of our various diffusion operators (kinematic viscosity, thermal conduction and dye concentration) follows the same algorithm described in \citet{may2006spectral}:
\begin{enumerate}
    \item Interpolate $U(x^s)\to U(x^f)$.
    \item Riemann problem: Choose left (or right).
    \item Compute the gradient $\nabla(U(x^f)) \to \nabla U(x^s)$.
    \item Interpolate $\nabla U(x^s)\to \nabla U(x^f)$.
    \item Riemann problem: Choose right (or left).
\end{enumerate}
For the SD method this can be achieve with the following operations. First, we used the operation described in \autoref{eq:sol} to perform the interpolation to flux points:
\begin{equation}
    \mathbf{U}(x^f_{a,m'}) = \sum_{m=0}^p \mathbf{U}(x^s_{a,m})\ell^s_m(x^f_{a,m'}).
\end{equation}

This solution is discontinuous at element boundaries. Since we will compute the gradient, we need to define a unique numerical value at the element boundaries. For this, we use the analogue of a Riemann solver $RP\{\mathbf{U}^L,\mathbf{U}^R\}$, so that the solution at flux points is now continuous and given by:
\begin{equation}
    \hat{\mathbf{U}}^f_{a,m} = 
    \begin{cases}
    RP\{\mathbf{U}^f_{a-1,p+1},\mathbf{U}^f_{a,0}\} \quad &m=0\\
    \mathbf{U}^f_{a,m} \quad &1\leq m\leq p \\
    RP\{\mathbf{U}^f_{a,p+1},\mathbf{U}^f_{a+1,0}\} \quad &m=p+1.
    \end{cases}
\end{equation} 

As there is no upwind direction in the parabolic case, as proposed by \citet{may2006spectral}, it suffices to choose one of the 2 options. For step (ii), we choose the left state:
\begin{equation}
    RP\{\mathbf{U}^L,\mathbf{U}^R\}= \mathbf{U}^L,
\end{equation}
and for step (v), we will choose the right state.\\

The third step involves a derivative from flux points to solution points (similar to the operator used in \autoref{eq:sd_update}), and therefore, requires the use the $\ell'$ polynomial basis (the derivative of the Lagrange polynomials):
\begin{equation}
     \nabla \mathbf{U}(x^s_{a,m}) = \nabla( \mathbf{U}(x^f_{a})) = \sum_{m'=0}^{p+1} \mathbf{U}(x^f_{a,m'}) \ell^{\prime f}_{m'}(x^s_{a,m}).
\end{equation}

The following step requires us to interpolate this gradient back to flux points, which can be performed similarly to \autoref{eq:sol}:
\begin{equation}
    \nabla\mathbf{U}(x^f_{a,m'}) = \sum_{m=0}^p \nabla\mathbf{U}(x^s_{a,m})\ell^s_m(x^f_{a,m'}).
\end{equation}

The final step consists of solving the Riemann problem for $\nabla\mathbf{U}(x^f_{a,m'})$, which, as mentioned before, should compensate the choice made for $\mathbf{U}(x^f_{a,m'})$. Therefore, the final step is given by enforcing:
\begin{equation}
    RP\{\nabla\mathbf{U}^L,\nabla\mathbf{U}^R\}= \nabla\mathbf{U}^R.
\end{equation}

At first order ($p=0$), this algorithm is equivalent to the standard three-point operator for the Laplacian, which is a second-order accurate method. We use this for our fall-back scheme in conjunction with FV2. For higher order ($p>1$), as we will show in the test section, the method is of order $p+1$. 

\section{Numerical results}

In this section we present a series of tests in 1, 2 and 3-dimensions to validate and asses the performance of the method previously described. In the following, 
we refer to the number of degrees of freedom $N_\text{DOF}$ as the number of cells for FV2 and the number of elements times the order [$N(p+1)$] for SD. This is in order to make a comparison as fair as possible between the different methods used here.
\label{sec:results}

\subsection{2D Taylor-Green vortex}

We use the 2-dimensional Taylor-Green vortex test to assess the convergence rates of the method when dealing with a rotating flow in the presence of dissipation, in the present case kinematic viscosity. The initial conditions for this test are as follows:
\begin{equation*}
\begin{aligned}
\rho &= \rho_0,\\
v_x &=  U_0 \sin(kx) \cos(ky),\\
v_y &= -U_0 \cos(kx) \sin(ky),\\
P &= P_0 - \frac{\rho U_0^2}{4} \left( \cos(2kx) + \cos(2ky)\right),
\end{aligned}
\end{equation*}
where $\rho_0=1$ is the initial mass density, $U_0$ is the initial characteristic velocity scale (initial amplitude), $k=1$ is the wavenumber (which determines the spatial scale of the vortex), and
$p_0=(\rho_0  (U_0 / \mathcal{M}_0)^2) / \gamma$ is the reference pressure, with $\mathcal{M}_0=0.1$ being the initial value for the Mach number and $\gamma=1.4$ the adiabatic index. The simulation domain is a box with $x \in [-\pi L,\pi L]$ and $y \in [-\pi L,\pi L]$ where $L=1/k$. The boundary conditions for this test are periodic in both directions.\\

\begin{figure}
    \centering
    \includegraphics[width=.9
    \linewidth]{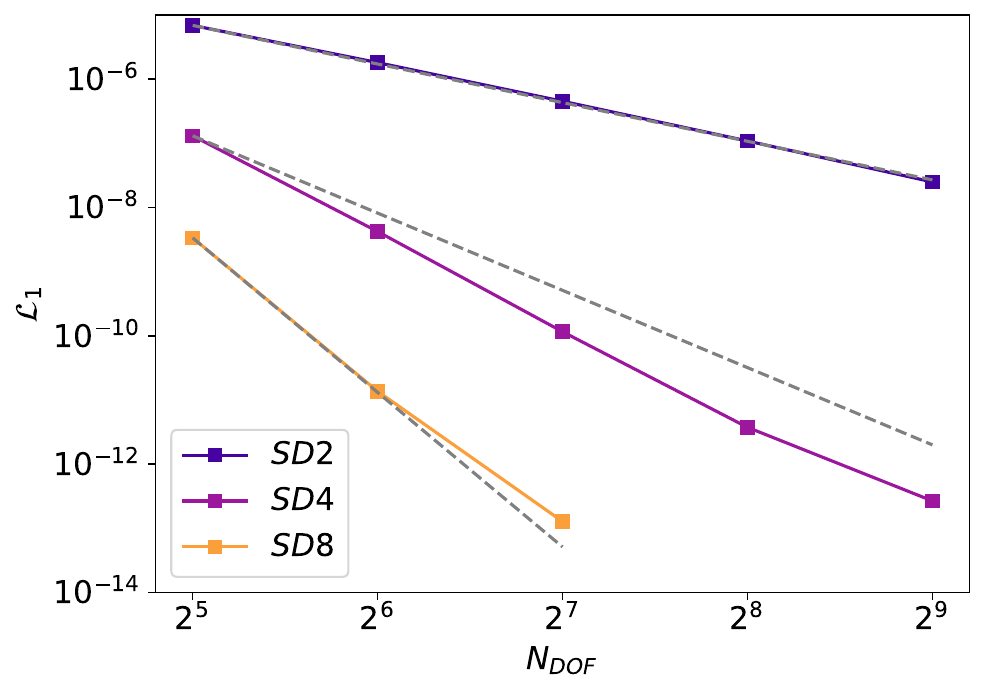}
    \caption{Convergence rates for the 2-dimensional viscous Taylor-Green vortex.}
    \label{fig:2dtgv}
\end{figure}

In 2D, the Taylor-Green vortex has a time-dependent analytical solution, which reads as follows:
\begin{equation*}
\begin{aligned}
v_x(t) &=  U_0 \sin(kx) \cos(ky) \exp(-2\nu t),\\
v_y(t) &= -U_0 \cos(kx) \sin(ky) \exp(-2\nu t),\\
P(t) &= P_0 - \frac{\rho U_0^2}{4} \left( \cos(2kx) + \cos(2ky)\right) \exp(-4\nu t),
\end{aligned}
\end{equation*}
where $\nu=0.01$ is the physical viscosity.
In \autoref{fig:2dtgv} we show the convergence rate of the method for three different orders of the numerical approximation (2nd, 4th, and 8th). The results are obtained by computing the $\mathcal{L}_1$ norm error at $t=1$ between the analytical and numerical solutions. One can observe that our method exhibits the desired convergence rates for this test. 
\subsection{Kelvin-Helmholtz instability}
We use the test described by \citet{lecoanet2016}.
\begin{equation}
     \rho = 1 + \frac{\Delta \rho}{\rho_0} \frac{1}{2} \left[ \tanh \left( \frac{y - y_1}{a} \right) - \tanh \left( \frac{y}{a} \right) \right]
\end{equation}
\begin{equation}
    v_x = v_{\text{flow}} \left[ \tanh \left( \frac{y - y_1}{a} \right) - \tanh \left( \frac{y}{a} \right) - 1 \right]
\end{equation}
\begin{equation}
    v_y = A \sin(2\pi y) \left[ \exp \left(-\frac{(y - y_1)^2}{\sigma^2} \right) + \exp \left(-\frac{(y - y_2)^2}{\sigma^2} \right) \right] 
\end{equation}
\begin{equation}
    P = P_0, \quad
    c = \frac{1}{2} \left[ \tanh \left( \frac{y - y_2}{a} \right) - \tanh \left( \frac{y - y_1}{a} \right) + 2 \right], 
\end{equation}
where $a = 0.05$, $\sigma = 0.2$, A = 0.01,$v_{\text{flow}}= 1$ and $p_0 = 10$. This results in a subsonic flow with Mach numbers $\mathcal{M}=0.25$ for $\rho = 1$ and $\mathcal{M}=0.35$ for $\rho = 2$. The domain is a rectangular box with $x \in [0, L]$ and $y \in [0, 2L]$, with $L = 1$ and $y_1 = 0.5$, $y_2 = 1.5$, with periodic boundary conditions in both directions.\\

We will be comparing the results of five different methods; FV2 (MUSCL-Hancock method), SD4 and SD8 (spectral difference at 4th and 8th order, respectively) and their limited counterparts SDB4 and SDB8.\\

\subsubsection{Uniform density}

We start with the case $\Delta \rho/\rho =0$, which corresponds to uniform density. We perform this test with and without limiting. With limiting, $c$ is the only variable used to detect troubled subcells.\\

\begin{figure}
    \centering
    \includegraphics[width=\linewidth]{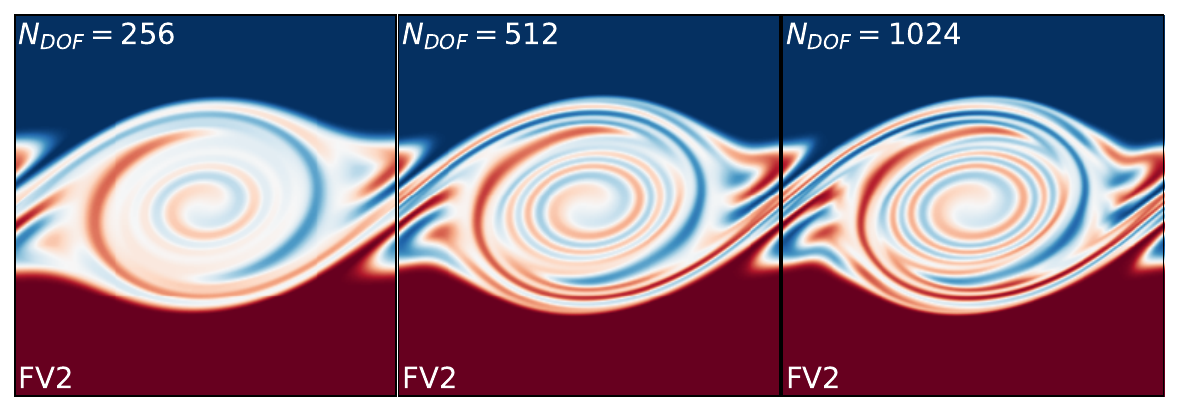}
    \caption{Color-maps for the dye concentration $C$ obtained with the FV2 (MUSCL-Hancock) method for the KHI test at $t=6$ with $\Delta\rho/\rho=0$ and $Re=10^5$.}
    \label{fig:drho0_fv}
\end{figure}

\begin{figure}
    \centering
    \includegraphics[width=1\linewidth]{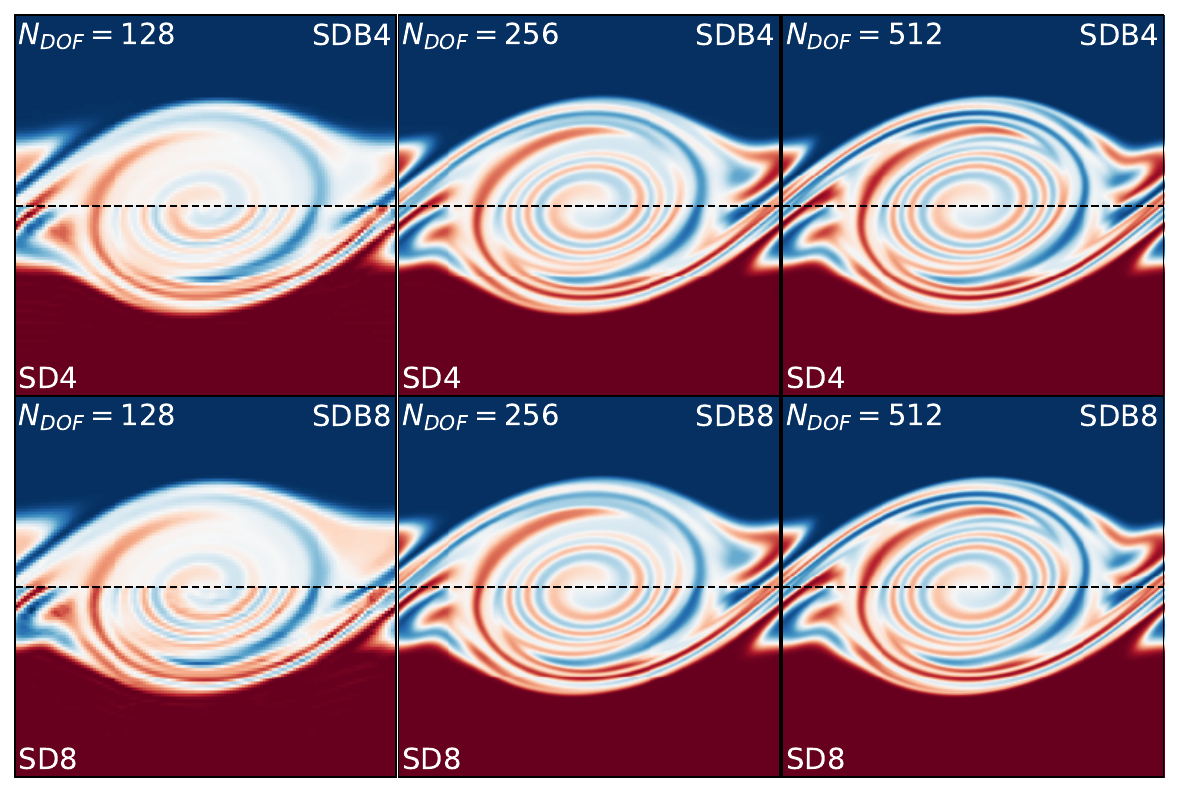}
    \caption{Color-maps for the dye concentration $C$ obtained with the SD method at 4th and 8th order for the KHI test at $t=6$ with $\Delta\rho/\rho=0$ $Re=10^5$. The upper half of each color-map shows the solution obtained when correcting for numerical oscillations (limiting), whereas the bottom half shows the solution without limiting.}
    \label{fig:drho0_sd}
\end{figure}

We first consider the case of a Reynolds number $Re = 10^5$. In \autoref{fig:drho0_fv} we present the results obtained for $t=6$ with our FV2 method for resolutions of $N_\text{DOF} = 256, 512$ and $1024$ (in the x-direction, two-fold in the y-direction). 
In \autoref{fig:drho0_sd} we present the results for $t=6$ for SD with 4th and 8th order (with and without limiting) with resolutions of $N_\text{DOF} = 128, 256$ and $512$. In the upper half of each color-map, we present the result with limiting, while in the lower half we present the solution without limiting. One can see that as the resolution is increased, and therefore the viscous scales start to be properly solved, the difference between limiting and non-limiting the solution vanishes.

\begin{figure*}
    \centering
    \includegraphics[width=1\linewidth]{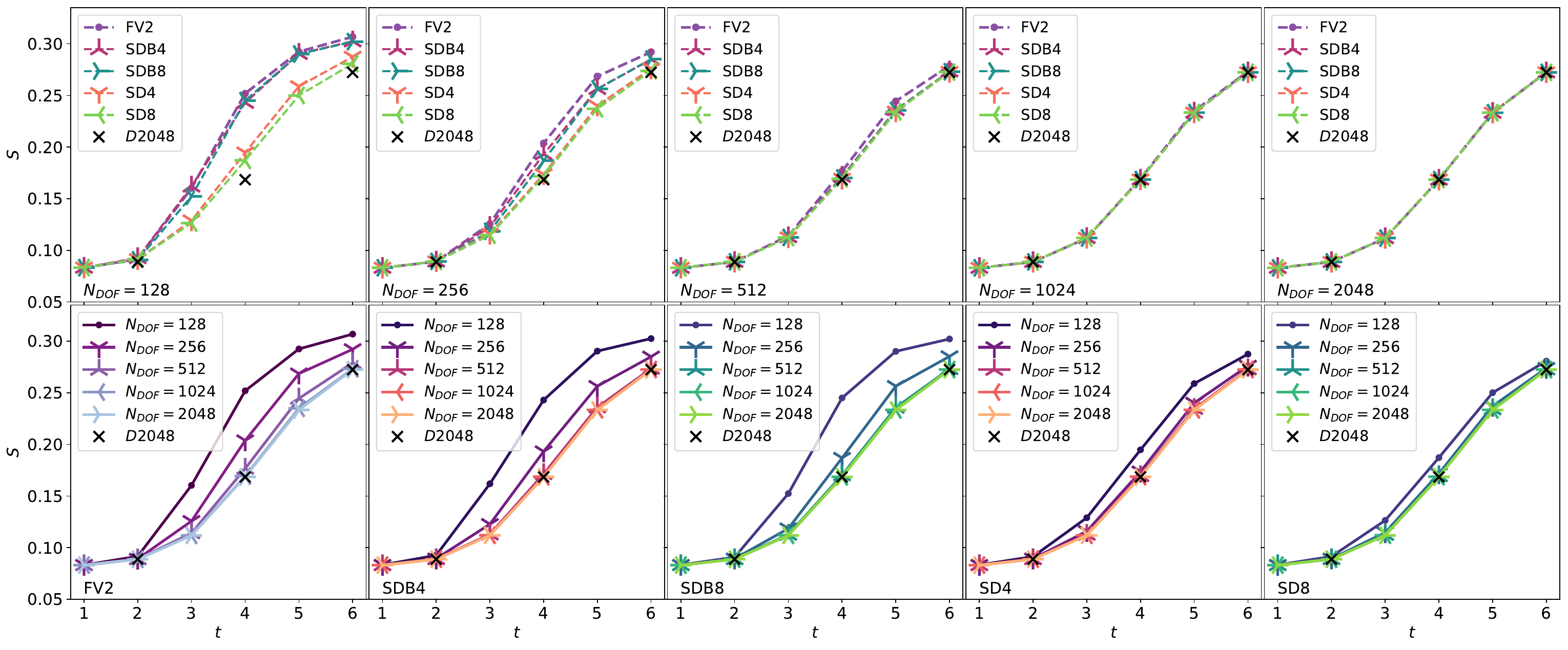}
    \caption{Entropy for the dye concentration $C$ as a function of time for the KHI test with $\Delta\rho/\rho=0$ and $Re=10^5$. We present results for the 5 methods used in this work, and as a reference the results obtained with \texttt{Dedalus}. On the upper row a comparison of the methods at a given resolution. On the lower row a comparison of resolutions for a given method.}
    \label{fig:entropy}
\end{figure*}

To quantify the performance of a given method, we monitored the evolution of the volume-integrated entropy $S$ for the dye concentration $c$:
\begin{equation}
    S = -\int_V \rho c \ln(c) dV,
\end{equation}
as described in \citet{lecoanet2016}. In \autoref{fig:entropy} we present the entropy $S$ as a function of time, measured for the five methods used in this work. In the upper row, we present the results for all methods at a given resolution. One can observe that at low resolution the results obtained by SD with limiting are only slightly better than the results obtained by FV2. As the resolution increases, the difference becomes more noticeable and the solution tends towards the unlimited high-order solution. In the bottom row, we present the same results arranged by method, comparing the results for different resolutions. This helps to observe the transition of the SDB4/8 from an FV2 dominated solution to and SD4/8 dominated one as the resolution is increased.\\

\begin{figure}
    \centering
    \includegraphics[width=1\linewidth]{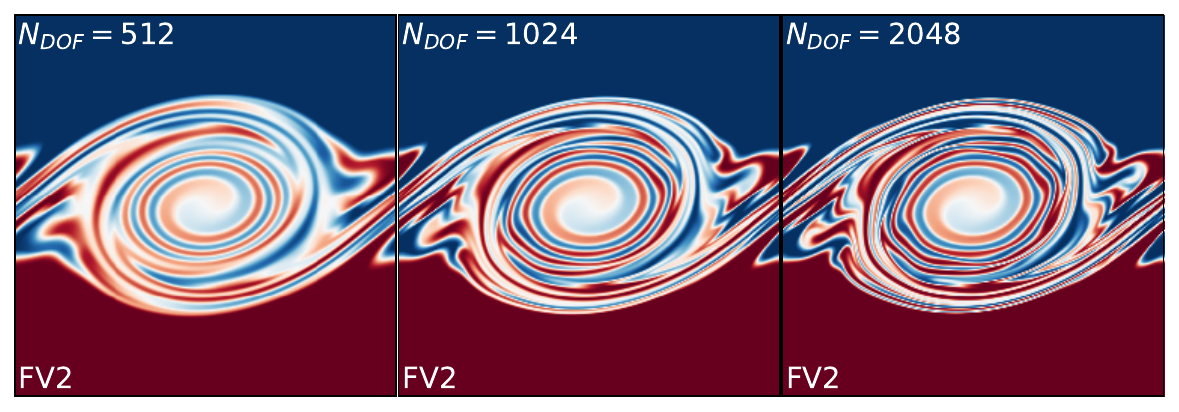}
    \caption{Color-maps for the dye concentration $C$ obtained with the FV2 (MUSCL-Hancock) method for the KHI test at $t=6$ with $\Delta\rho/\rho=0$ and $Re=10^6$.}
    \label{fig:drho0_fv_Re6}
\end{figure}

\begin{figure}
    \centering
    \includegraphics[width=1\linewidth]{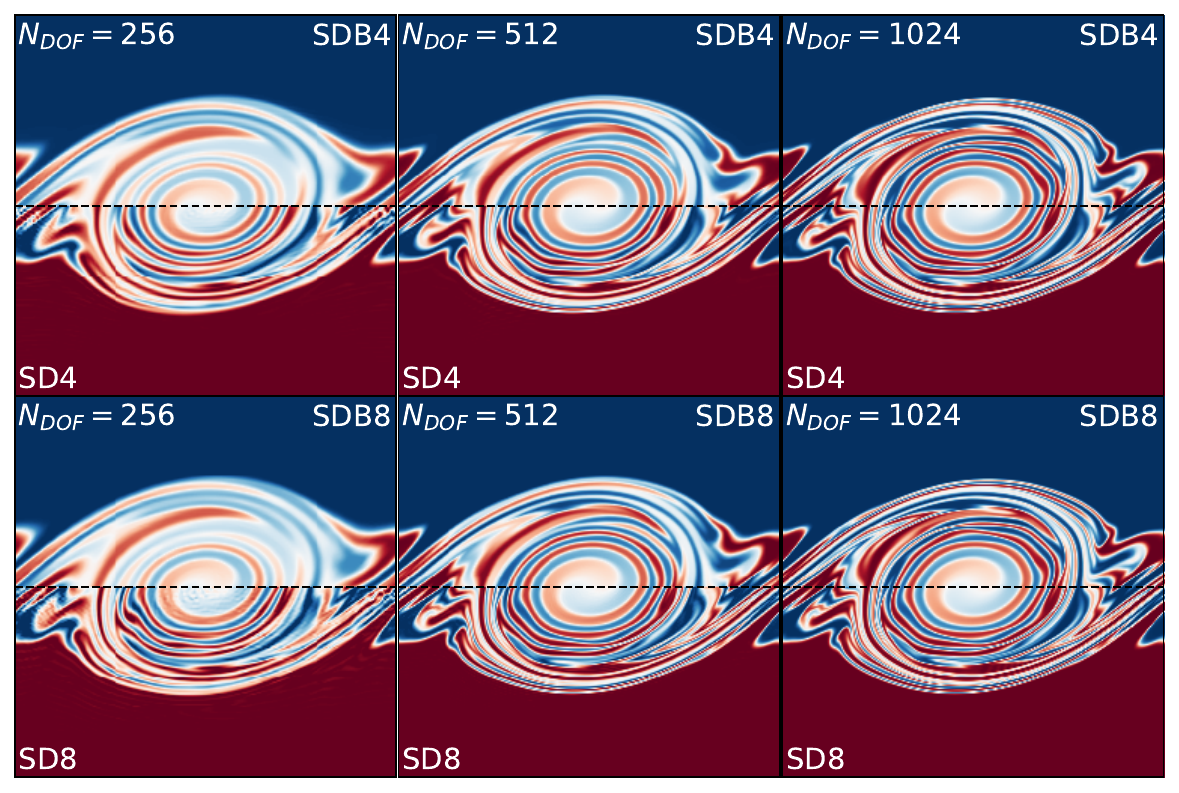}
    \caption{Color-maps for the dye concentration $C$ obtained with the SD method at 4th and 8th order for the KHI test at $t=6$ with $\Delta\rho/\rho=0$ $Re=10^6$. The upper half of each color-map shows the solution obtained when correcting for numerical oscillations (limiting), whereas the bottom half shows the solution without limiting.}
    \label{fig:drho0_sd_Re6}
\end{figure}

We continue our analysis with a test for a Reynolds number of $Re = 10^6$ which results in a more non-linear solution. In \autoref{fig:drho0_fv_Re6} we present the results for FV2 at $t=6$ with $N_\text{DOF} = 512, 1024$ and $2048$.
In \autoref{fig:drho0_sd_Re6} we present the results for SD at $t=6$ with $N_\text{DOF} = 256, 512$ and $1024$, half the number of degrees of freedom used with FV2. We observe a similar trend as for $Re=10^5$, with an FV2 dominated solution at low resolution and an SD dominated solution as the resolution increases.\\

We continue by comparing the evolution of the entropy for the dye concentration $S$ as a function of time. This should help to elucidate the impact of limiting at low resolution, as well as the impact of high-order as the resolution is increased. 

\begin{figure*}
    \centering
    \includegraphics[width=1\linewidth]{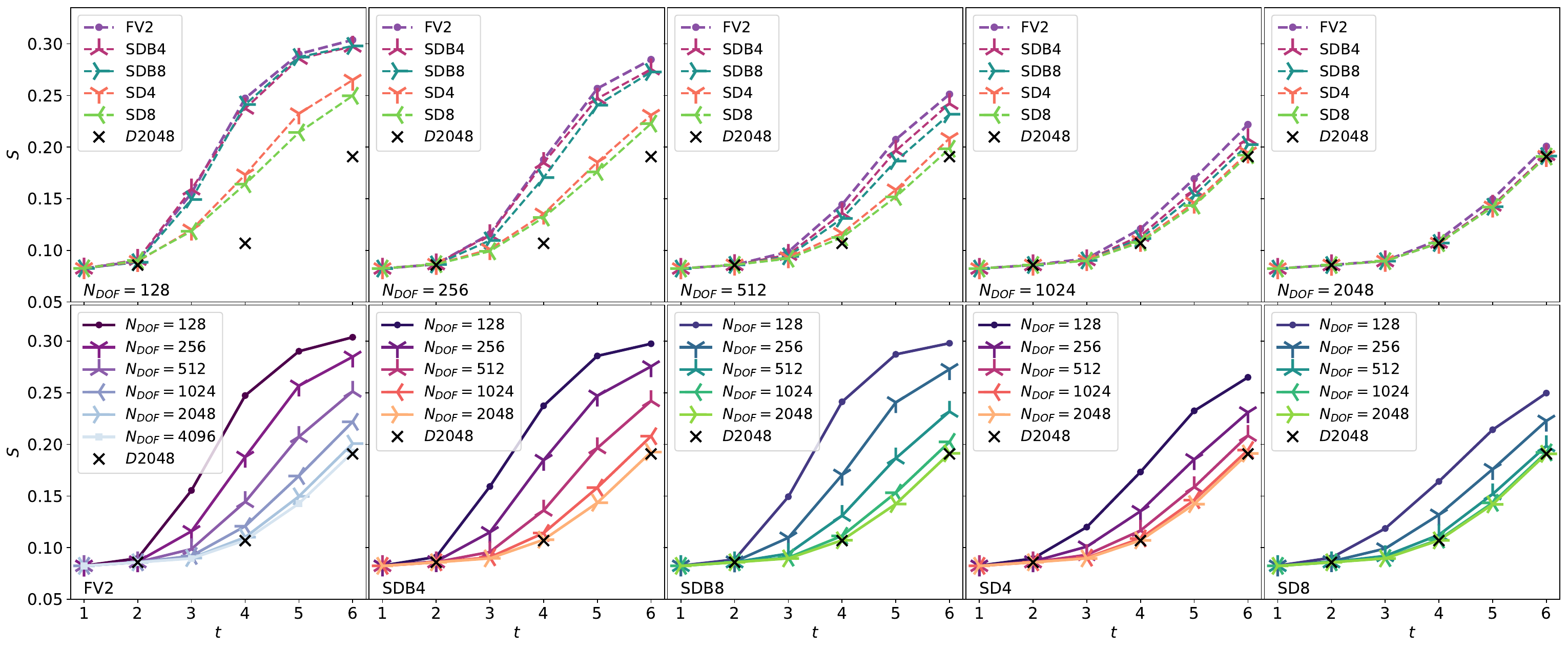}
    \caption{Entropy for the dye concentration $C$ as a function of time for the KHI test with $\Delta\rho/\rho=0$ and $Re=10^6$. We present results for the 5 methods used in this work, and as a reference the results obtained with \texttt{Dedalus}. On the upper row a comparison of the methods at a given resolution. On the lower row a comparison of resolutions for a given method.}
    \label{fig:entropy_Re6}
\end{figure*}

In \autoref{fig:entropy_Re6} we present the equivalent of \autoref{fig:entropy} for $Re=10^6$. We observe a faster convergence at high-order, seeing converged results at $N_\text{DOF}=2048$ for SD in the four variants used here.

\subsubsection{Convergence}

In \autoref{tab:Re5} we present the convergence rates for the $\mathcal{L}_2$ norm error when running the KHI test for $Re=10^5$. 
\begin{table}
    \centering
    \begin{tabular}{|c|cccccc}
      p/t & $1$ & $2$ & $3$ &$4$ & $5$ & $6$\\
      \hline
      $1$ & 2.23 & 1.88 & 1.9 & 1.82 & 2.09 & 2.22\\
      $3$ & 3.96 & 4.56 & 4.55 & 4.57 & 4.47 & 4.82\\
      $7$ & 6.34 & 6.64 & 7.07 & 7.45 & 6.85 & 6.61\\
    \end{tabular}
    \caption{Convergence rates for the $\mathcal{L}_2$ obtained with SD for the KHI test with $Re=10^5$}
    \label{tab:Re5}
\end{table}
\autoref{tab:Re6} presents the results for the $Re=10^6$ case. One can clearly observe that the convergence rate is affected by the Reynolds number, deteriorating as it increases. This trend is due to the more chaotic nature of the flow as the viscosity vanishes, creating more small-scale structures.
\begin{table}
    \centering
    \begin{tabular}{|c|cccccc}
      p/t & $1$ & $2$ & $3$ &$4$ & $5$ & $6$\\
      \hline
      $1$ & 2.04 & 1.29 & 1.02 & 1.0 & 0.85 &0.93\\
      $3$ & 4.0 & 2.55 & 1.98 & 1.99 & 1.94 & 2.07\\
      $7$ & 4.63 & 4.85 & 3.81 & 3.7 & 3.78 & 3.66\\
    \end{tabular}
    \caption{Convergence rates for the $\mathcal{L}_2$ obtained with SD for the KHI test with $Re=10^6$}
    \label{tab:Re6}
\end{table}

\subsubsection{Stratified case}

We consider now the same test, but this time with $\Delta \rho/\rho = 1$ which results in a jump in density. We perform the test for a Reynolds number of $Re=10^5$. In this case the detection of troubled cells is performed using $\rho$, $P$ and $c$, not just $c$.\\

\begin{figure}
    \centering
    \includegraphics[width=1\linewidth]{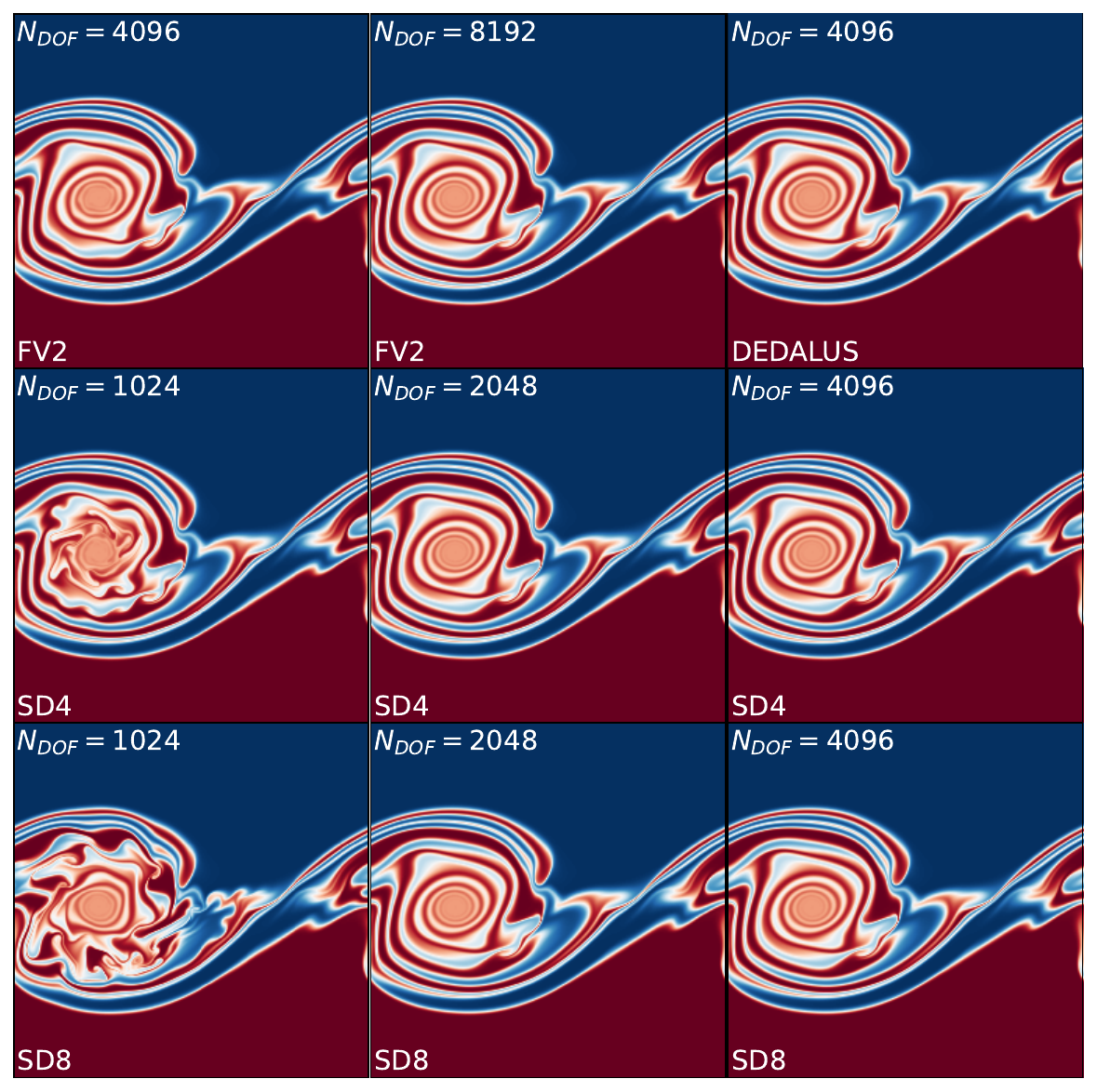}
    \caption{Color-maps for the dye concentration $C$ obtained with FV2 and the SD method at 4th and 8th order for the KHI test at $t=4$ with $\Delta\rho/\rho=1$ and $Re=10^5$. The upper row shows, in the first two columns, the results for FV2 (MUSCL-Hancock) at $N_\text{DOF}=4096$ and $8192$, and for comparison, in the last column, the results for \texttt{Dedalus} at $N_\text{DOF}=4096$. The middle and bottom row show the results for SD4 and SD8 (respectively) at $N_\text{DOF}=1024,\ 2048$ and $4096$. At this time ($t=4$), the results for SD4 and SD8 appear to be converged at $N_\text{DOF}=2048$, whereas for FV2 convergence is achieved at $N_\text{DOF}=8096$.}
    \label{fig:drho1_t4}
\end{figure}

In the first row of \autoref{fig:drho1_t4}, we show color-maps for the results obtained at $t=4$ for the FV2 method. One can see that even at $N_\text{DOF}=4096$ the inner spiral presents small disturbances, in contrast to the converged solution of \citet{lecoanet2016} (shown in the third column). The second and third rows in \autoref{fig:drho1_t4} show the results for SD and $t=4$. In this case, we can observe the converged solution of \citet{lecoanet2016, Athena++} even at $N_\text{DOF}=2048$, with no evidence of differences between the limited and unlimited solutions.\\

\begin{figure}
    \centering
    \includegraphics[width=1\linewidth]{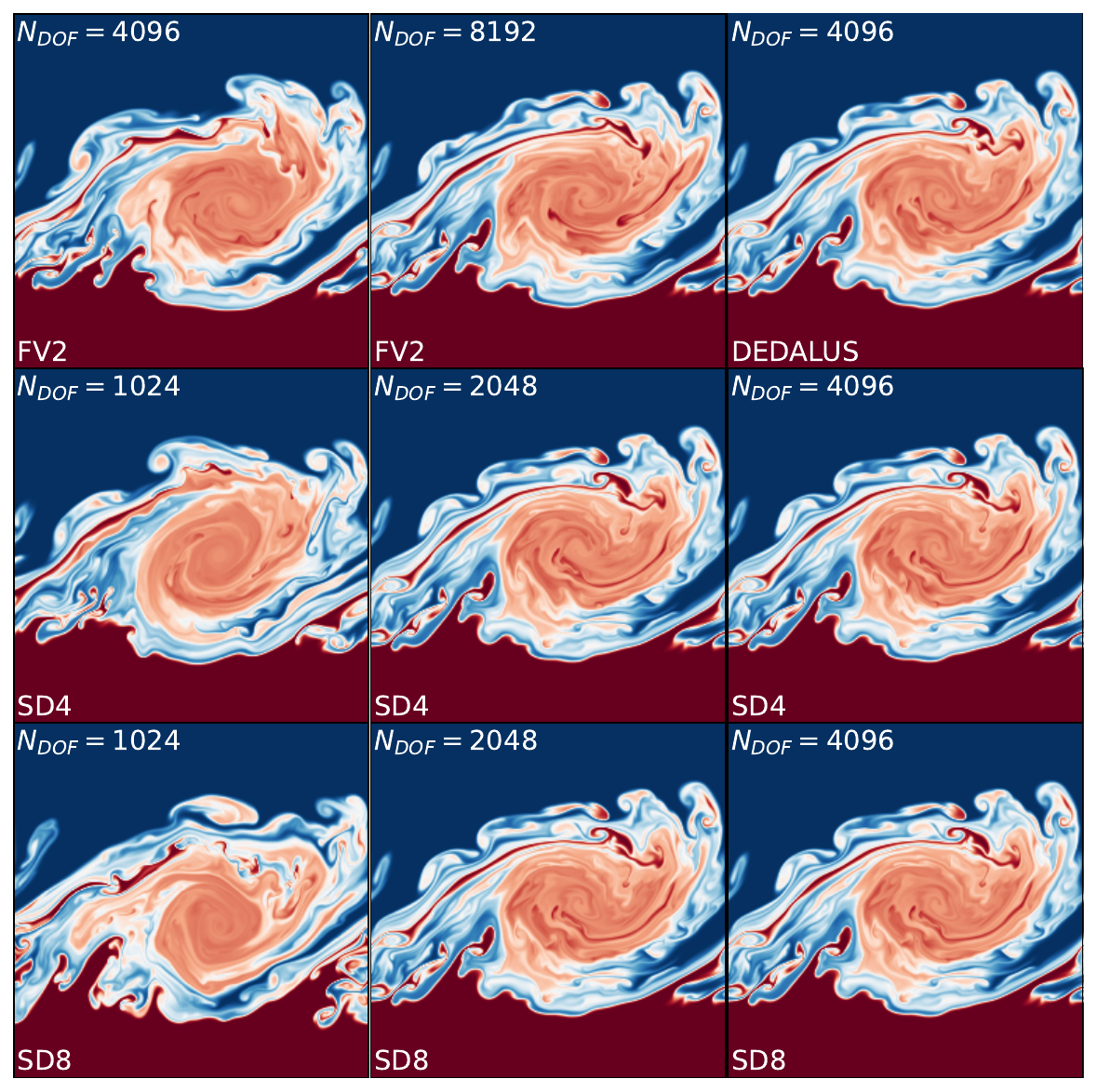}
    \caption{Color-maps for the dye concentration $C$ obtained with FV2 and the SD method at 4th and 8th order for the KHI test at $t=6$ with $\Delta\rho/\rho=1$ and $Re=10^5$. The upper row shows, in the first two columns, the results for FV2 (MUSCL-Hancock) at $N_\text{DOF}=4096$ and $8192$, and for comparison, in the last column, the results for \texttt{Dedalus} at $N_\text{DOF}=4096$. The middle and bottom row show the results for SD4 and SD8 (respectively) at $N_\text{DOF}=1024,\ 2048$ and $4096$. At this time ($t=6$), the results for SD4 reach convergence at $N_\text{DOF}=4096$, and for SD8 at $N_\text{DOF}=2048$. while for FV2, convergence is not achieved even at $N_\text{DOF}=8096$.}
    \label{fig:drho1_t6}
\end{figure}

In the first row of \autoref{fig:drho1_t6} we show the results for the FV2 method and $t=6$. We include the results for $N_\text{DOF}= 4096$ and $8192$, where even at the highest resolution the solution is far from resembling the  converged solution of \citet{lecoanet2016,Athena++} (shown in the third column). In the second and third rows of \autoref{fig:drho1_t6} we show the results for SD (and $t=6$) with  $N_\text{DOF}= 1024, 2048$ and $4096$, where we can see that the external structure has reached convergence even at $N_\text{DOF}= 2048$ for both 4th and 8th order, even with limiting. We can see that the internal structure closely resembles the converged solution obtained by \citet{lecoanet2016, Athena++}, with the results at 8th order and $N_\text{DOF}= 2048$ being quite similar to those at 4th order $N_\text{DOF}= 4096$.

\subsection{3-dimensional Taylor-Green vortex}

\begin{figure}
    \centering
    \includegraphics[width=1.2\linewidth]{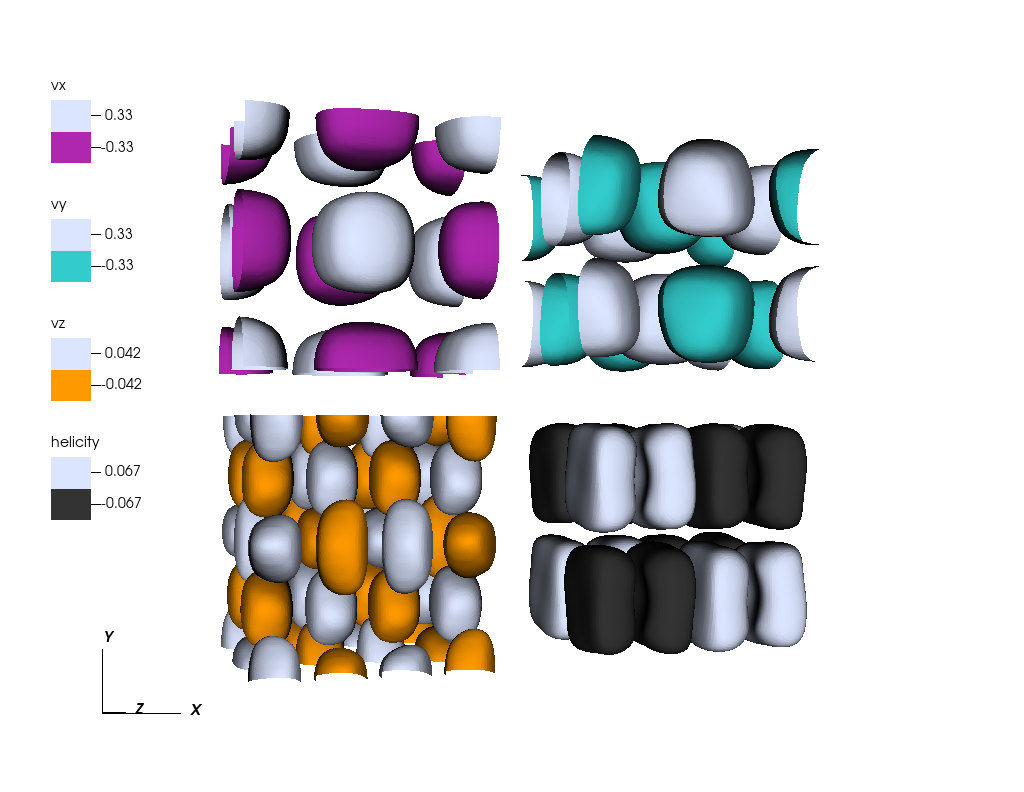}
    \caption{Iso-surfaces of the velocity components and helicity for the 3-dimensional Taylor-Green vortex at $t=0.5$ for $Re=800$ and $\mathcal{M}_0=0.1$, obtained with SD$8$ and $N_{DOF}=128^3$.}
    \label{fig:3DTGV-t05}
\end{figure}

\begin{figure}
    \centering
    \includegraphics[width=1.2\linewidth]{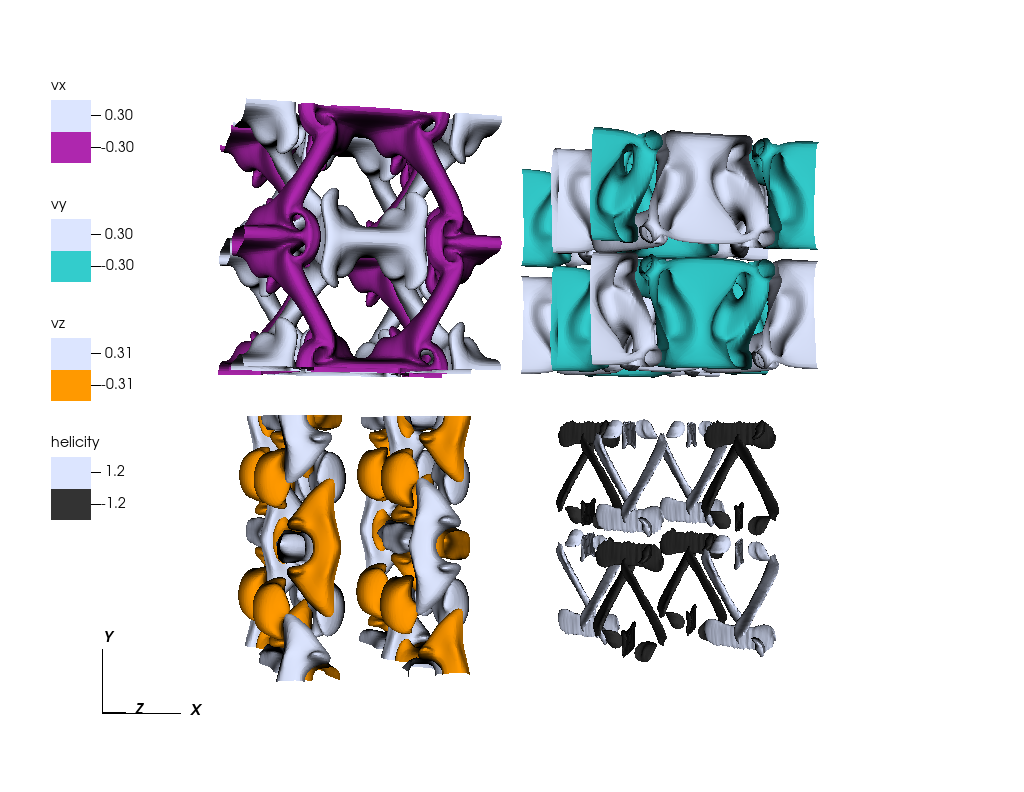}
    \caption{Iso-surfaces of the velocity components and helicity for the 3-dimensional Taylor-Green vortex at $t=5$ for $Re=800$ and $\mathcal{M}_0=0.1$, obtained with SD$8$ and $N_{DOF}=128^3$.}
    \label{fig:3DTGV-t5}
\end{figure}

\begin{figure}
    \centering
    \includegraphics[width=1.2\linewidth]{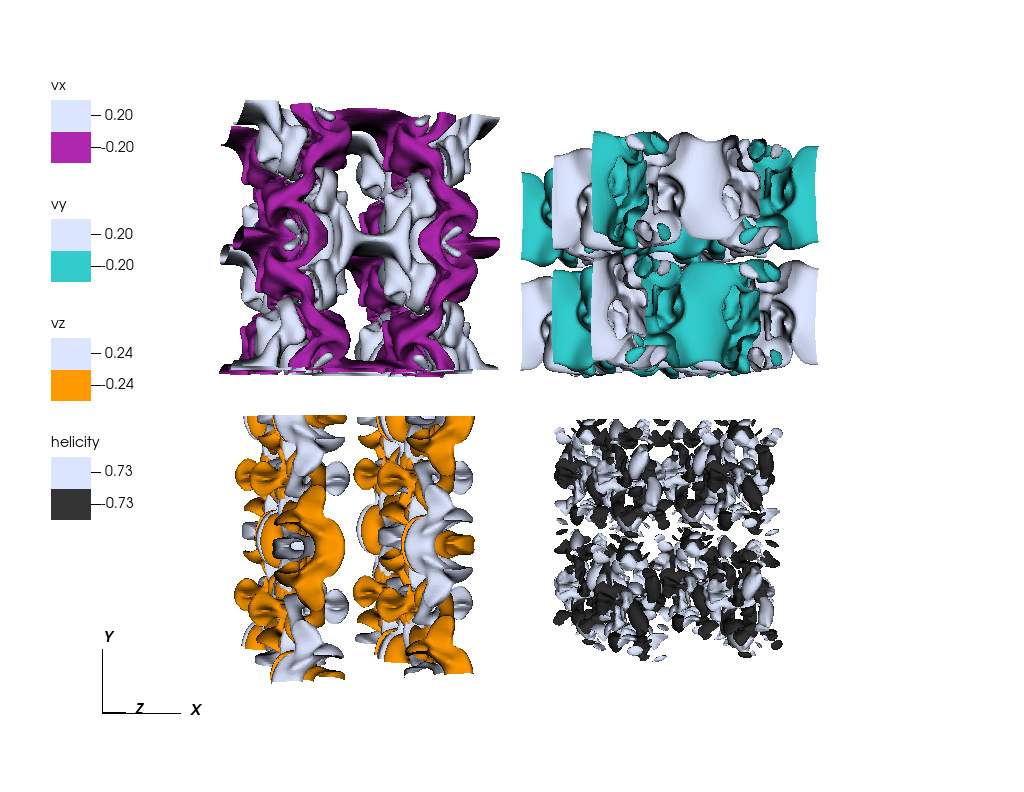}
    \caption{Iso-surfaces of the velocity components and helicity for the 3-dimensional Taylor-Green vortex at $t=10$ for $Re=800$ and $\mathcal{M}_0=0.1$, obtained with SD$8$ and $N_{DOF}=128^3$.}
    \label{fig:3DTGV-t10}
\end{figure}

The initial conditions for the 3D Taylor-Green vortex \citep{brachet1983small,orszag2005numerical} are defined as follows:
\begin{equation*}
\begin{aligned}
\rho &= 1\\
v_x &= U_0 \sin(kx) \cos(ky) \cos(kz) \\
v_y &= -U_0 \cos(kx) \sin(ky) \cos(kz) \\
v_z &= 0 \\
P &= P_0 - \frac{\rho U_0^2}{4} \left( \cos(2kx) + \cos(2ky) + \cos(2kz) \right)
\end{aligned}
\end{equation*}
where $U_0$ is the characteristic velocity scale (the initial amplitude), $k=1$ is the wavenumber (which determines the spatial scale of the vortex), $\nu=0.01$ is the physical viscosity, and
$p_0=(\rho_0  (U_0 / \mathcal{M}_0)^2) / \gamma$ is the reference pressure, with $\mathcal{M}_0=0.1$ being the initial value for the Mach number and $\gamma=1.4$ the adiabatic index. The simulation domain is a box with $x \in [-\pi L,\pi L]$, $y \in [-\pi L,\pi L]$, and $z \in [-\pi L,\pi L]$ where $L=1/k$. The boundary conditions for this test are periodic in either direction.
We start with a simulation for $Re=800$, using SD$8$ and $N_{DOF}=128^3$.
In \autoref{fig:3DTGV-t05} we show iso-surfaces at $t=0.5$ for the three components of the velocity field, as well as for the helicity. At this early time we can observe the stretching of the vortex. \autoref{fig:3DTGV-t5} follows the same simulation up to $t=5$, where we can see the vortex roll-up phase. Finally, \autoref{fig:3DTGV-t10} shows the vortex breaking down at $t=10$, showing the early onset of turbulence. 

\begin{figure}
    \centering
    \includegraphics[width=.9\linewidth]{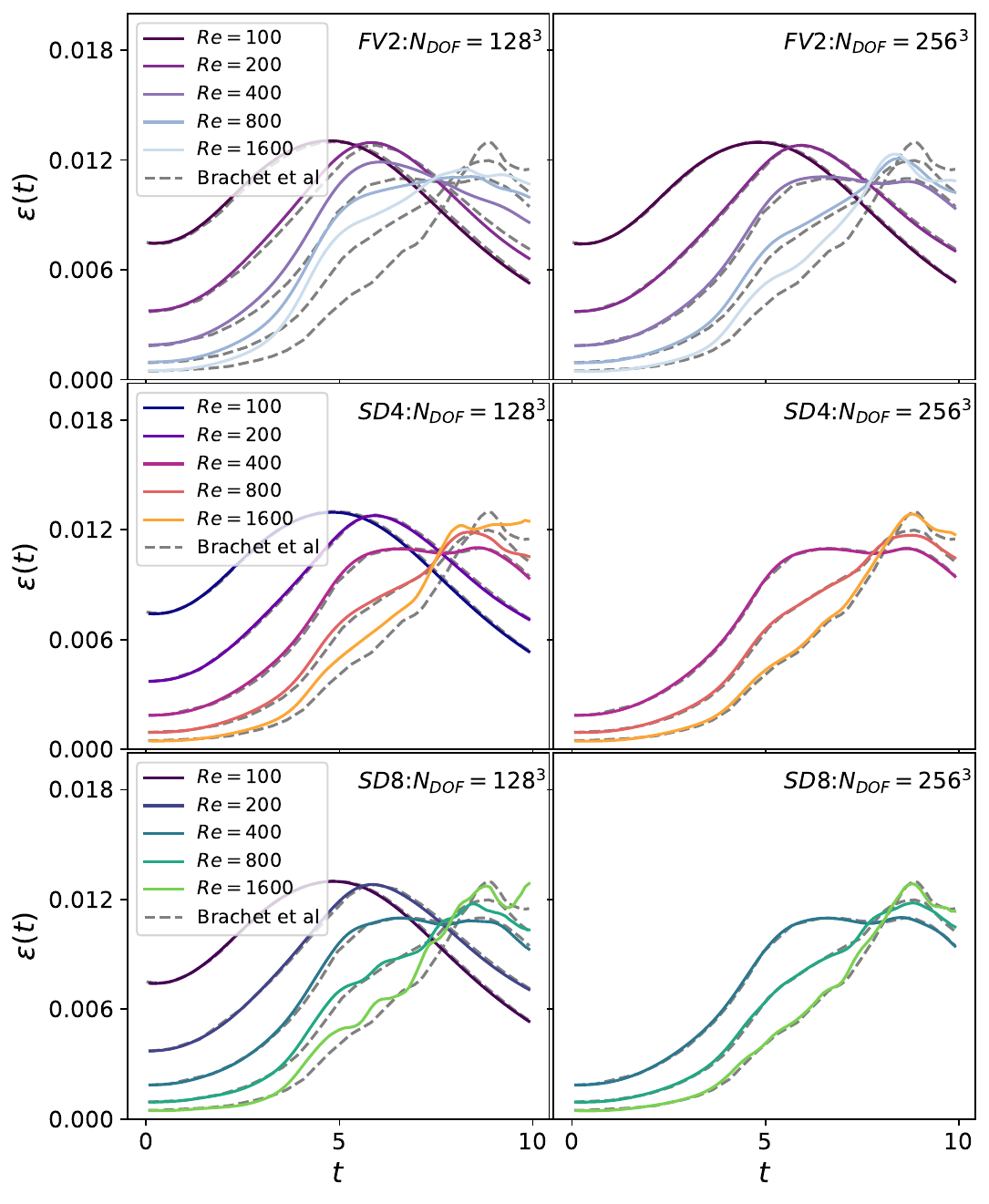}
    \caption{Dissipation rate for the kinetic energy as a function of time $\epsilon(t)$ , computed for the 3-dimensional Taylor-Green vortex at different values of the Reynolds number ($Re=100,200,400,800$) and for a Mach number $\mathcal{M}_0=0.1$ (therefore a subsonic flow). The mosaic shows the results for FV2, SD4 and SD8 with $N_{DOF}=128^3$ and $256^3$.}
    \label{fig:3dtgv}
\end{figure}
We ran simulations for different values of the Reynolds number ($Re=100,200,400,800$ and $1600$) similar to the study of \citet{brachet1983small}. We compute the dissipation rate $\epsilon$ for the kinetic energy $E_k$, and monitored this as a function of time.
\begin{equation}
    \epsilon = \frac{dE_k}{dt} = \frac{d}{dt}\left( \frac{1}{|V|}\int_{V} \frac{\rho}{2} \boldsymbol{v} \cdot \boldsymbol{v} dV\right).
\end{equation}
In \autoref{fig:3dtgv} we compare the results for FV2, SD4 and SD8 using $N_{DOF}=128^3$ and $256^3$. We can observe that at $N_{DOF}=128^3$, SD4 and SD8 are able to replicate the converged results of \citet{brachet1983small} for $Re=400$, while FV2 achieves this at $N_{DOF}=256^3$. Moreover, at $N_{DOF}=256^3$, SD4 and SD8 are able to capture the results for $Re=800$ and almost those for $Re=1600$. This results highlight the benefits of using high-order methods when dealing with smooth flows.

\section{Discussion}
\label{sec:discussion}
In this work, we have presented a spectral difference (SD) method incorporating an arbitrarily high-order treatment of the Laplacian operator and a posteriori limiting. Our results demonstrate that this approach effectively captures both smooth features and discontinuities in fluid flows while accurately resolving diffusive scales, even at lower resolutions than traditional lower-order methods. Moreover, the a posteriori limiting ensures robustness, preventing instabilities even when viscous scales are not fully resolved.

Similar studies have been conducted at high order for the Navier-Stokes equations. We start comparing our results with those obtained by \citet{lecoanet2016} for their well-posed problem a KH instability. Said work includes a comparison between different methods. First \texttt{DEDALUS} \citep{burns2020dedalus}, a code based on the Fourier spectral method, therefore, providing high-order solutions, with the caveat that it does not possess shock capture capabilities. Secondly, \texttt{Athena++} \citep{Athena++}, an FV method based on a fourth-order implementation of the piecewise parabolic method (PPM) \citep{colella1984piecewise}. Note that both methods used a third-order Runge Kutta method for the time integration. 

For the unstratified case and a Reynolds number of $Re=10^5$, they show that both \texttt{DEDALUS} and \texttt{ATHENA++} exhibit convergence at $N_\text{DOF}=1024$, with the error for \texttt{DEDALUS} being several orders of magnitude smaller. They also show that at low resolution, the error for \texttt{DEDALUS} is dominated by time integration, showing that by reducing the time step, they can achieve converged results with only $N_\text{DOF}=512$, similar to our results. 

When moving to the stratified case, the difference between the methods is clearly enhanced in favor of \texttt{DEDALUS} (without taking into account the computational cost). In this case, \texttt{DEDALUS} shows converged results at $N_\text{DOF}=2048$ while \texttt{ATHENA++} needs eight-folds that to achieve the same. This is in accordance with our results, where we observe convergence for SD$8$ at $N_\text{DOF}=2048$, and for SD$4$ at $N_\text{DOF}=4096$. We wish to stress here that our method combines the best of these two methods, resolving the viscous scales at lower resolution (like \texttt{DEDALUS}), and being robust when that is not the case (like \texttt{ATHENA++}). 

In \citet{Athena++} a follow-up of the previous comparison is shown including two different reconstruction methods; PLM (Piecewise linear method), a 2nd order method \citep{van1977towards}, and the 4th order PPM. Finding that the results for \citet{Athena++} and PPM in \citet{lecoanet2016}, were dominated by the numerical diffusion of the Laplacian operator, being implemented at only 2nd order. Showing that with a high-order version of this operator, \citet{Athena++} can achieve convergence for the stratified KHI test at $N_\text{DOF}=8192$ for PPM (4th order) which is twice the resolution for which our SD4 achieves convergence. 

We continue by comparing our results with those of the DG method with limiting through artificial viscosity described in \citet{2023MNRAS.522..982C}. Similar to us, their high-order implementation exhibits exponential convergence at the proper rate when dealing with smooth solutions. A study of the KHI test of \citet{lecoanet2016} is presented for
$\Delta\rho/\rho=0$ and $R=10^5$ with $64$ elements and $p=2, 4$ and $5$ (third, fifth, and sixth order). This amounts to resolutions of $N_\text{DOF} = 192, 320$ and $384$ respectively, where the simulation at sixth order and $N_\text{DOF} = 384$, is the only one that has reached convergence. These results agree with our own, where we observe convergence only at $N_\text{DOF} = 512$ with 4th and 8th order. 


Other high-order methods have been investigated in their ability to describe the 3-dimensional Taylor-Green vortex at different Reynolds numbers. In \citet{tavelli2016staggered} they present a staggered space-time DG method for unstructured tetrahedral meshes. This method is able to closely match the results of \citet{brachet1983small} up to $Re=1600$ by using $494592$ elements of polynomial degree $p=4$. In \citet{boscheri2021high} a method based on CWENO (central weighted essentially non-oscillatory) reconstruction \citep{levy1999central} and semi-implicit IMEX (implicit explicit) time integration \citep{boscarino2019high}, achieving third order of accuracy is tested up to $Re=400$ making use of $120^3$ cells. These results agree with our own, being outperformed by our 4th and 8th order implementations at a slightly higher resolution ($N_\text{DOF}=128^3$). \citet{fehn2021high} present results at $Re=1600$ for their ALE-DG (arbitrarily Lagrangian Eulerian DG) method making use of up to $N_\text{DOF}=128^3$ for 4th order and a moving mesh, obtaining results slightly better than ours at the same resolution and order. \citet{diosady2015case} present results for a DG implementation making use of 2th, 4th, 8th and 16th order in space and 4th order in time, with up to $N_\text{DOF}=256^3$ for $Re=1600$. Their results are in good agreement with ours, showing a similar trend when increasing the order. 

\section{Conclusions}
\label{sec:conclusions}
We have implemented an arbitrarily high-order method for the diffusion (Laplacian) operator for kinematic viscosity, thermal conduction and dye concentration diffusivity into the SD framework. Our findings build upon previous work based on the SD method [\paperI, \paperII] by explicitly addressing the challenge of accurately representing these diffusion terms at high order. 

We have tested the convergence of the resulting method when dealing with smooth solutions, in 1-dimension for the diffusion of a sine wave, and in 2-dimensions for the Taylor-Green vortex, obtaining the expected exponential convergence. 

We have also tested the performance of the method in the presence of discontinuities, namely the KHI test of \citet{lecoanet2016}. We have shown that the results obtained with SD$8$ are comparable to those obtained with \texttt{DEDALUS} with the same number of degrees of freedom, and the results with SD$4$ outperform those of \texttt{Athena++} with piecewise parabolic reconstruction (PPM). Note that we have not compared to computational cost of these different methods, which is an important aspect when deciding what method to use.

We have also showed that our SD method is capable of performing well, even when the viscous scale is not properly solved, unlike classical spectral and pseudo-spectral methods who usually fail. We have also showed that the higher order SD schemes (SD4 and SD8) are capable of resolving these scales at lower grid resolution when compared to lower order methods.

In light of the presence study, we believe that high order finite element methods in general, and the SD method in particular, are very powerful schemes to solve for smooth solutions typical of subsonic and viscous flows. SD turned out to be particularly appealing as an accurate (exponential convergence) but also robust (limiting) method in this case.

\section*{Acknowledgements}

The simulations included in this work were executed on the Stellar cluster at Princeton University.

\section*{Data availability statement}
The data underlying this article will be shared on reasonable request to the corresponding author.


\bibliographystyle{mnras}
\bibliography{biblio} 



\appendix


\bsp	
\label{lastpage}
\end{document}